\documentclass[conference]{IEEEtran}

\usepackage[T1]{fontenc}
\usepackage[utf8]{inputenc}
\usepackage{amsmath,amssymb,amsfonts}
\usepackage{graphicx}
\usepackage{textcomp}
\usepackage{xcolor}
\usepackage{booktabs}
\usepackage{array}
\usepackage{url}

\usepackage{microtype}                
\usepackage[backend=biber, style=ieee, sorting=none]{biblatex}
\begin{document}

\title{AspisAI: A Canonical, Machine-Interpretable Governance Framework for Automated Multi-Standard Compliance Monitoring}

\author{%
\IEEEauthorblockN{Tsafac Nkombong Regine Cyrille}
\IEEEauthorblockA{\textit{CyberMACS, Applied Cybersecurity} \\ tsafacnkombong.cyrille@stu.khas.edu.tr}

\IEEEauthorblockN{Hasan Dag}
\IEEEauthorblockA{\textit{Kadir Has University} \\ Istanbul, T\"urkiye}

\IEEEauthorblockN{Reiner Creutzburg}
\IEEEauthorblockA{\textit{SRH University of Applied Sciences Heidelberg} \\ Campus Berlin, Germany}

\IEEEauthorblockN{Knut Haufe}
\IEEEauthorblockA{\textit{SRH University of Applied Sciences Heidelberg} \\ Campus Berlin, Germany}
}

\maketitle

% ============================================================
\begin{abstract}
Organisations operating in regulated and critical-infrastructure sectors must satisfy multiple, heterogeneous cybersecurity and privacy instruments simultaneously, including but not limited to ISO/IEC~27001, the NIST Cybersecurity Framework~2.0, Cyber Essentials, and the GDPR. In practice these obligations are managed through manual mappings, spreadsheet-based tracking, and periodic audits that are costly to maintain, inconsistent across standards, and weak in traceability. This paper presents \emph{AspisAI}, a bounded, standard-agnostic governance framework that translates selected requirements from several frameworks into a canonical, machine-interpretable control model, and evaluates submitted evidence against condition-based decision rules to produce explainable, traceable compliance determinations. Within a bounded scope of 26 representative requirements, the framework is evaluated in a controlled simulation against five governance-oriented criteria and, critically, against two external reference points that mitigate the circularity of single-author evaluation: its cross-standard mappings are validated against NIST's own published informative references, with 57\,\% exact agreement and divergences confined to same-family controls, and the framework is applied to real third-party evidence from the OpenSSF Scorecard, surfacing genuine governance gaps in a live open-source project. The controlled results, comprising full requirement encoding, 88.5\,\% mapping coverage, complete traceability, and correct detection of all introduced gaps, establish functional correctness, while the external validation provides evidence of applicability beyond the simulation. The contribution is therefore a demonstration that a canonical, provenance-preserving governance model can render multi-standard compliance both automatable and auditable.
\end{abstract}

\begin{IEEEkeywords}
Compliance automation, cybersecurity governance, machine-interpretable policy, ISO/IEC 27001, NIST CSF, GDPR, critical infrastructure, design science research, traceability.
\end{IEEEkeywords}

% ============================================================
\section{Introduction}
\IEEEPARstart{C}{ompliance} with multiple cybersecurity and privacy frameworks has become a routine obligation for organisations in regulated sectors. Instruments such as ISO/IEC~27001, the NIST Cybersecurity Framework~2.0, Cyber Essentials, and the GDPR overlap substantially in intent yet differ in structure, vocabulary, and evidentiary expectations \cite{wang2024survey, mubarkoot2023software}. Managing them together typically relies on static, manually maintained mappings that drift out of alignment with the underlying standards, duplicate evidence-collection effort, and provide weak traceability from a compliance outcome back to the evidence that supports it.

Emerging approaches such as compliance automation, policy-as-code, and machine-readable governance artefacts suggest that compliance logic can be formalised \cite{barhaim2023cloud, haverinen2024devsecops, angermeir2024automated}. However, much of this work is implementation-heavy or specific to particular technical environments such as cloud platforms or DevSecOps pipelines, and gives comparatively little attention to the governance-design problem itself: how selected requirements from heterogeneous frameworks can first be translated into a bounded, traceable, machine-interpretable governance structure.

This paper addresses that gap with \emph{AspisAI}, a governance framework that translates selected cybersecurity, privacy, and resilience requirements into structured, machine-interpretable governance logic suitable for automated and continuous compliance monitoring. The contributions are: (i)~a standard-agnostic canonical control model with accompanying JSON schemas that preserves the provenance of each requirement; (ii)~a lightweight rule-evaluation engine that produces explainable, traceable compliance determinations; and (iii)~an evaluation on a simulated critical-infrastructure case study against five governance-oriented criteria.

% ============================================================
\section{Related Work}
Three streams of prior work are relevant. The first compares cybersecurity and regulatory frameworks and documents their overlap and the difficulty of aligning them consistently in practice \cite{wang2024survey, mubarkoot2023software}. The second concerns compliance automation: Bar-Haim \textit{et al.} assess organisational cybersecurity posture in cloud settings using structured representations \cite{barhaim2023cloud}; Haverinen \textit{et al.} automate compliance in DevSecOps through an open information model for security-as-code \cite{haverinen2024devsecops}; and Angermeir \textit{et al.} examine continuous security compliance \cite{angermeir2024automated}. The third addresses privacy- and GDPR-oriented compliance \cite{aberkane2021gdpr, nist_privacy, klymenko2022technical, amaral2023nlp} and the heightened need for traceable governance in critical-infrastructure contexts \cite{nis2, ruohonen2026systematicliteraturereviewnis2}.

A fourth, emerging stream treats compliance as executable artefacts. Policy-as-code expresses governance constraints directly as machine-checkable code \cite{haverinen2024devsecops}, and recent work pairs large language models with deterministic validators so that machine-generated mappings or policies are checked before being trusted \cite{Romeo_2025, schoning2026complianceaisystems, marino2024compliancecardsautomatedeu}. This stream is methodologically closest to the present work, which adopts the same ``propose-then-validate'' discipline but anchors it in a canonical cross-standard model rather than a single policy language.

Across these streams, prior work tends to satisfy one or two of the properties required for integrated governance, namely multi-standard scope, machine-interpretability, and end-to-end traceability, but rarely all three within a single artefact. Frameworks that automate checking within a single environment, for example cloud posture or a CI/CD pipeline, typically lack a standard-agnostic canonical layer, while cross-framework comparison studies establish where standards overlap but stop short of an executable, traceable artefact. AspisAI is positioned to address this combination explicitly, with emphasis on a canonical model that preserves provenance while supporting cross-standard mapping and interpretable evaluation. This is best characterised as a \emph{technical gap}: existing approaches are individually capable but are not composed into a single traceable, multi-standard governance layer.

% ============================================================
\section{Methodology}
\begin{figure}[!t]
\centering
\includegraphics[width=\linewidth]{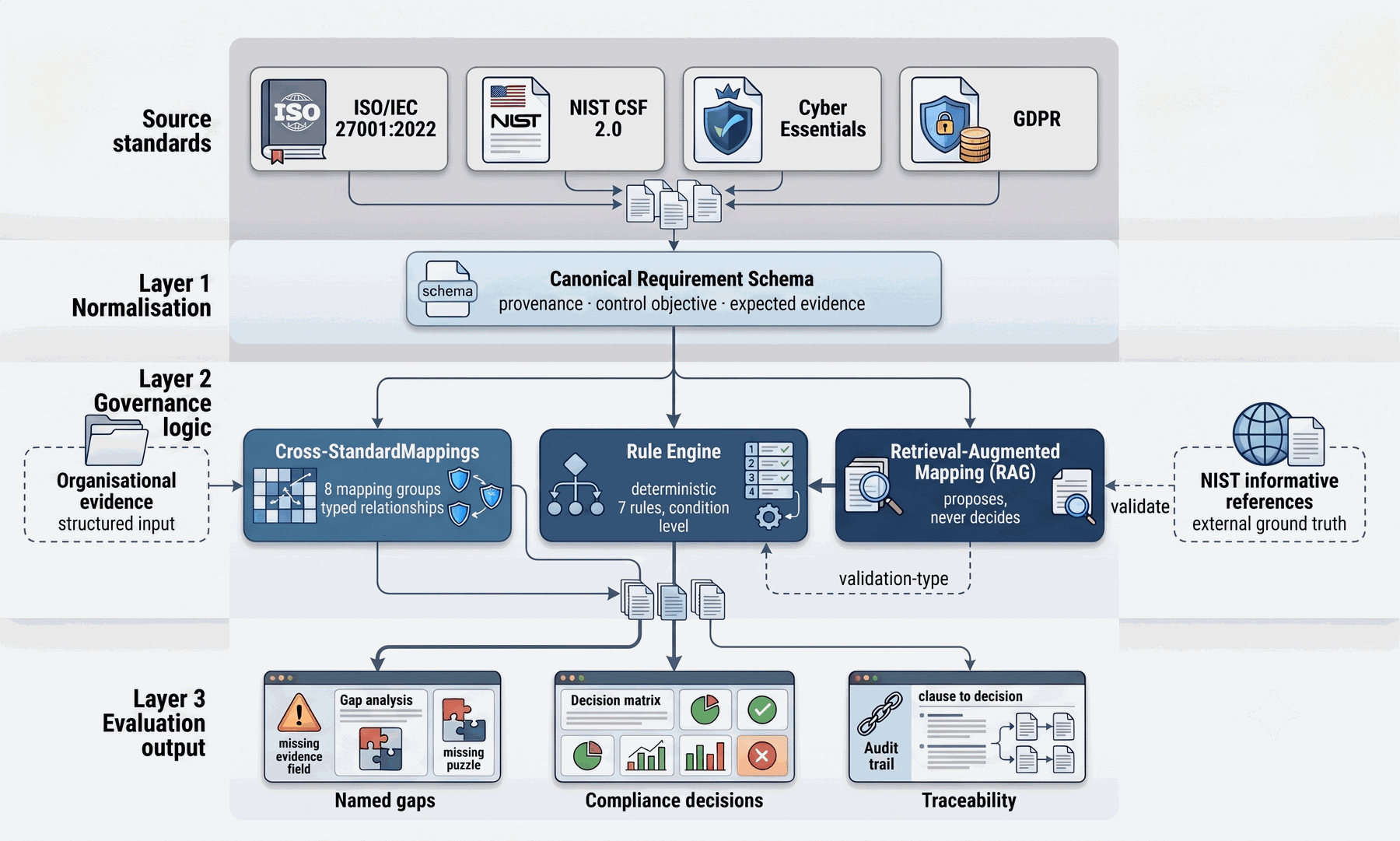}
\caption{The three-layer AspisAI architecture. Layer~1 normalises requirements from four standards into a single canonical schema; Layer~2 applies cross-standard mapping and the rule engine to submitted evidence; Layer~3 produces traceable compliance decisions and gap reports.}
\label{fig:arch}
\end{figure}

The study follows a design science research (DSR) methodology \cite{peffers2007dsrm, hevner2004design}, organised around the design and evaluation of an artefact. The artefact comprises five tightly related components: a bounded set of selected requirements, a canonical control structure, cross-standard mappings, machine-interpretable governance rules, and traceability-preserving compliance outputs. Figure~\ref{fig:arch} shows how these components compose into three layers: normalisation, governance logic, and evaluation output.

\subsection{Canonical Schema and Requirement Selection}
A bounded set of 26 representative requirements was selected from ISO/IEC 27001, NIST CSF~2.0, Cyber Essentials, and the GDPR. Each requirement is encoded in a canonical JSON schema with mandatory fields including a unique identifier, source standard, clause reference, control objective, governance domain, and expected evidence. This common representation establishes the shared vocabulary needed for cross-standard mapping and preserves the provenance of every requirement. Table~\ref{tab:sources} summarises the four primary sources and the contextual instrument.

\begin{table}[!t]
\renewcommand{\arraystretch}{1.3}
\caption{Primary Requirement Sources in the Canonical Model}
\label{tab:sources}
\centering
\footnotesize
\setlength{\tabcolsep}{4pt}
\begin{tabular}{@{}l l@{}}
\toprule
\textbf{Source} & \textbf{Role in the model} \\
\midrule
ISO/IEC 27001:2022 & ISMS governance requirements \\
NIST CSF 2.0 & High-level cybersecurity outcomes \\
Cyber Essentials & Prescriptive technical baseline \\
GDPR (Arts.\ 5, 25, 32, 33, 35) & Privacy and data-protection duties \\
NIS2 Directive & Critical-infrastructure context \\
\bottomrule
\end{tabular}
\end{table}

\subsection{Governance Logic and Rule Engine}
A governance logic layer operates on the normalised representation through two components: a cross-standard mapping that records equivalence, overlap, and complementarity between requirements; and a rule engine that evaluates submitted evidence.

A rule $R$ is a finite set of conditions $C_R = \{c_1, \dots, c_n\}$, where each condition $c_i = (f_i, \mathit{op}_i, v_i)$ names an evidence field, a comparison operator, and an expected value. Submitted evidence $E$ is treated as a partial map from field names to observed values, formed by merging the property sets of every artefact in the submission. A condition is satisfied only if the field was actually supplied and the comparison holds:
\begin{equation}
\sigma_i =
\begin{cases}
1 & \text{if } f_i \in \mathrm{dom}(E) \text{ and } \mathit{op}_i(E(f_i), v_i)\\
0 & \text{otherwise}
\end{cases}
\label{eq:sat}
\end{equation}
Writing $k = \sum_{i=1}^{n} \sigma_i$ for the number of satisfied conditions, the determination is
\begin{equation}
D(R, E) =
\begin{cases}
\textsc{compliant} & \text{if } k = n\\
\textsc{partial} & \text{if } 0 < k < n\\
\textsc{non-compliant} & \text{if } k = 0
\end{cases}
\label{eq:determination}
\end{equation}
with compliance score $s = k/n$ and gap set $G(R,E) = \{f_i \mid \sigma_i = 0\}$. The gap set is what makes an output actionable: it names the specific fields that failed rather than returning a verdict alone. Because $\sigma_i$ requires $f_i \in \mathrm{dom}(E)$, a field that was never supplied is treated as unsatisfied rather than silently skipped, which is the mechanism by which missing evidence becomes a detectable gap.

The engine deliberately uses straightforward Boolean logic rather than weighted or probabilistic scoring to preserve interpretability, and binds every rule to one or more canonical requirements so that each determination is traceable to its source clause. Because the rule engine is purely Boolean and contains no stochastic element, the governance evaluation is fully deterministic: re-executing it on identical inputs yields byte-identical decisions, which makes the reported results reproducible from the accompanying artefact. Non-determinism is confined to the optional AI-assisted mapping component (Section~\ref{sec:aimapping}), and is deliberately contained by routing every model proposal through the deterministic validator before it can affect any decision.

\subsection{Simulated Case Study}
The framework is demonstrated on \emph{NorthGrid Energy Distribution Ltd}, a simulated regional energy utility with hybrid IT/OT infrastructure. Four synthetic evidence domains (access control, incident response, data protection, and risk assessment) were defined, with a number of intentional gaps introduced to test detection. Synthetic evidence avoids reliance on sensitive organisational data while preserving realism.

\subsection{Evaluation Criteria}
The artefact is evaluated against five governance-oriented criteria: coverage (proportion of selected requirements encoded), mapping quality (cross-standard correspondence), rule decision accuracy (agreement with a predefined ground truth), traceability completeness (unbroken requirement-to-evidence chains), and gap detection (correct identification of missing evidence).

\subsection{AI-Assisted Mapping with Deterministic Validation}
\label{sec:aimapping}
To explore how the canonical model could scale beyond manually curated mappings, an optional component allows a large language model to \emph{propose} cross-standard relationships, which the framework then \emph{validates} deterministically. A proposal is accepted only if both requirement identifiers exist in the canonical model, the proposed relationship type is permitted, and the relationship is consistent with NIST's published informative references; otherwise it is flagged as a probable hallucination. The language model is therefore never trusted directly: it acts as a generator of candidate mappings, while authority rests with the deterministic validator. Because the language model is itself non-deterministic, the component supports repeated execution so that the run-to-run stability of accepted mappings can be measured; the validator itself is deterministic and produces identical verdicts on identical proposals.

% ============================================================
\section{Results}
\label{sec:results}
\subsection{Coverage and Mapping Quality}
All 26 selected requirements were successfully encoded and validated against the canonical schema. The overall cross-standard mapping coverage rate, the proportion of requirements appearing in at least one mapping group, was 88.5\,\% (23 of 26). Coverage was not uniform across frameworks; the three unmapped requirements were source-specific controls without natural cross-standard counterparts, reflecting realistic correspondence rather than forced linkage.

\subsection{Decision Accuracy and Gap Detection}
Eleven case-study evaluations were performed. Eight returned \emph{compliant} (72.7\,\%) and three returned \emph{partially compliant} (27.3\,\%), consistent with the intentional evidence design. Against a predefined ground truth, the engine matched the expected status in all 11 cases (100\,\% ground-truth accuracy). All three intentionally introduced gaps were correctly identified (100\,\% gap detection), with each output record naming the specific missing evidence property.

\subsection{Traceability}
\begin{figure*}[!t]
\centering
\includegraphics[width=0.92\textwidth]{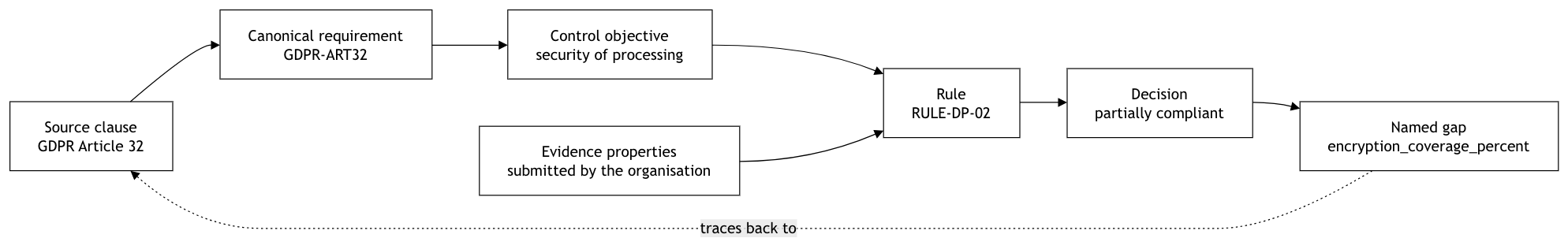}
\caption{The traceability chain retained by every determination, from the source clause through the canonical requirement and the applied rule to the evidence property evaluated and any gap named. Each link is stored, so a determination can be reconstructed after the fact by a party who was not present when it ran.}
\label{fig:trace}
\end{figure*}

Every compliance determination retained an unbroken chain from the source clause, through the canonical requirement and the bound rule, to the specific evidence property evaluated (Figure~\ref{fig:trace}). The chain is walkable in both directions, which is what distinguishes an output usable in an audit from a bare verdict: instead of reporting non-compliance, the system names the property that fell short and the clause requiring it. This supports audit and accountability functions that statistical security-monitoring systems do not provide.

% Spans both columns: this diagram is wide and is unreadable at column width.

\subsection{A Worked Determination}
\label{sec:worked}
One evaluation makes the mechanism concrete. Rule \texttt{RULE-AC-01} is bound to three requirements drawn from three different standards: ISO/IEC~27001 Annex~A~8.5 (secure authentication), NIST CSF~2.0 \texttt{PR.AA-01}, and Cyber Essentials \texttt{CE-AC-01}. Its condition set is $C_R = \{(\textit{policy\_approved},=,\textsc{true}),\ (\textit{mfa\_enabled},=,\textsc{true}),\ (\textit{mfa\_coverage\_percent},\geq,90)\}$.

The submitted evidence \texttt{EV-AC-001} bundles three artefacts, an approved access control policy, an MFA configuration report, and an access log sample. Merging their properties yields $\textit{policy\_approved} = \textsc{true}$, $\textit{mfa\_enabled} = \textsc{true}$, and $\textit{mfa\_coverage\_percent} = 95$. All three conditions hold, so $k = n = 3$, and by~\eqref{eq:determination} the determination is \textsc{compliant} with $s = 1.0$ and $G = \emptyset$ (Table~\ref{tab:worked}).

The significant point is what happened once: a single evidence submission was evaluated a single time, and it discharged obligations under three separate standards. The engine records this as three determinations, one per bound requirement, each carrying its own clause reference so that an ISO auditor and a Cyber Essentials assessor can each be answered from the same underlying evaluation. Removing that duplication is the practical case for a canonical layer.

The contrasting case shows how a gap surfaces. Rule \texttt{RULE-DP-02}, bound to GDPR Article~32, requires encryption at rest, encryption in transit, and coverage of at least 95\,\%. The evidence reports the first two as true but coverage at 85\,\%, one database having been left pending an upgrade. Here $k = 2$, $n = 3$, so the determination is \textsc{partial} with $s = 0.667$ and $G = \{\textit{encryption\_coverage\_percent}\}$. The output names the field, the observed value, and the threshold it missed, which is the difference between a finding an engineer can act on and a score.

\begin{table}[!t]
\renewcommand{\arraystretch}{1.3}
\caption{A Worked Determination for Rule \texttt{RULE-AC-01}}
\label{tab:worked}
\centering
\footnotesize
\setlength{\tabcolsep}{4pt}
\begin{tabular}{@{}l p{5.1cm}@{}}
\toprule
\textbf{Element} & \textbf{Value} \\
\midrule
Bound requirements & ISO/IEC 27001 A.8.5; NIST CSF PR.AA-01; CE-AC-01 \\
Conditions $C_R$ & \textit{policy\_approved} $=$ true; \textit{mfa\_enabled} $=$ true; \textit{mfa\_coverage\_percent} $\geq 90$ \\
Evidence & \texttt{EV-AC-001} (3 artefacts) \\
Observed values & true; true; 95 \\
Satisfied $k/n$ & 3/3 \\
Determination & \textsc{compliant}, $s = 1.0$ \\
Gap set $G$ & $\emptyset$ \\
\bottomrule
\end{tabular}
\end{table}

\subsection{External Validation}
Two checks address the circularity inherent in evaluating a single-author artefact against its own ground truth. First, the framework's CSF and ISO mappings were compared against NIST's official informative references: 4 of 7 asserted mappings (57.1\,\%) matched exactly, with the three divergences resolving to controls within the same ISO family rather than to genuine errors (Table~\ref{tab:mapval}).

\begin{table}[!t]
\renewcommand{\arraystretch}{1.3}
\caption{Mapping Validation Against NIST Informative References}
\label{tab:mapval}
\centering
\footnotesize
\setlength{\tabcolsep}{4pt}
\begin{tabular}{@{}l c@{}}
\toprule
\textbf{Outcome} & \textbf{Count} \\
\midrule
Exact agreement with NIST references & 4/7 (57.1\,\%) \\
Same-family divergence (not an error) & 3/7 \\
Genuine mismatch & 0/7 \\
\bottomrule
\end{tabular}
\end{table}

Second, the framework was applied to real, externally generated evidence from the OpenSSF Scorecard for the \emph{Prometheus} project, where it correctly identified 8 of 10 controls as compliant and surfaced two genuine governance gaps: unresolved dependency vulnerabilities and unsigned releases. Neither reference was authored by the researcher, so agreement in the first case and gap discovery in the second provide evidence of validity that the controlled case study alone cannot.

\subsection{Baseline Comparison}
To isolate the contribution of the canonical model, the framework was compared against two simpler baselines. A keyword-matching mapper, which links requirements by lexical overlap alone, recovered the curated cross-standard mappings poorly (F1 $\approx$ 0.30), confirming that naive text similarity cannot substitute for a curated canonical structure. A rule-only configuration, which evaluates evidence without the canonical mapping layer, achieved zero cross-standard traceability and produced no cross-standard mapping groups, against complete traceability and eight mapping groups for the full framework. These comparisons attribute the framework's traceability and cross-standard capabilities specifically to the canonical model rather than to the rule engine alone. Table~\ref{tab:baseline} summarises the comparison.

% Four columns do not fit one IEEE column; spans both.
\begin{table*}[!t]
\renewcommand{\arraystretch}{1.3}
\caption{Framework versus Simpler Baselines}
\label{tab:baseline}
\centering
\begin{tabular}{l c c c}
\toprule
\textbf{Approach} & \textbf{Mapping F1} & \textbf{Traceability} & \textbf{Mapping groups} \\
\midrule
Keyword matching & 0.30 & n/a & n/a \\
Rule-only (no canonical model) & n/a & 0.0 & 0 \\
\textbf{AspisAI (full)} & \textbf{curated} & \textbf{1.0} & \textbf{8} \\
\bottomrule
\end{tabular}
\end{table*}

\subsection{Adversarial Robustness}
A set of adversarial probes was applied to test the evaluation logic beyond its intended inputs. Two genuine limitations were surfaced: the engine does not currently distinguish \emph{absent} evidence from \emph{failing} evidence in numeric comparisons, and it accepts evidence that is \emph{declared} but not substantiated. These findings bound the framework's claims to \emph{declared} compliance and motivate evidence authentication as future work; importantly, they were discovered through the evaluation rather than assumed away.

Table~\ref{tab:summary} collects the results of all five criteria together with the two external checks.

\begin{table}[!t]
\renewcommand{\arraystretch}{1.3}
\caption{Summary of Evaluation Results (Simulation Scope)}
\label{tab:summary}
\centering
\footnotesize
\setlength{\tabcolsep}{4pt}
\begin{tabular}{@{}l c@{}}
\toprule
\textbf{Criterion} & \textbf{Result} \\
\midrule
Requirement coverage (encoded) & 26/26 (100\,\%) \\
Cross-standard mapping coverage & 23/26 (88.5\,\%) \\
Rule decision accuracy (ground truth) & 11/11 (100\,\%) \\
Traceability completeness & Complete \\
Gap detection & 3/3 (100\,\%) \\
Mapping validity (vs.\ NIST refs) & 4/7 (57.1\,\%) \\
Real-evidence case (OpenSSF) & 8/10, 2 gaps found \\
\bottomrule
\end{tabular}
\end{table}

% ============================================================
\section{Discussion}
The results support a specific claim: that the governance-design problem, translating heterogeneous requirements into a bounded, traceable, machine-interpretable model, can be solved in a way that is both automatable and externally checkable. This is distinct from, and prior to, the engineering problem of building a compliance scanner; the canonical model is the contribution, and the rule engine merely demonstrates that it is executable. Within the bounded scope, the lightweight canonical layer consolidated multi-standard requirements into a single interpretable model, automated repetitive checks, and surfaced cross-standard gaps while preserving traceability. The principal limitations are the bounded requirement set (26 items), the use of a simulated case study rather than production data, and a condition-based rule language that trades expressiveness for interpretability. These constrain external validity but are appropriate for establishing feasibility and design soundness.

The single-author construction of the case study, in which rules, evidence, and ground truth share one author, means the controlled results should be read as evidence of \emph{functional correctness and internal consistency} rather than of generalisation. This is precisely why the external checks of Section~\ref{sec:results}, mapping validation against NIST references and application to real OpenSSF evidence, carry the weight of the validity argument: they are the components an independent party did not author.

\subsection{Threats to Validity}
Four categories of threat apply, and each is stated with the mitigation actually taken rather than the one that would have been ideal.

\textit{Construct validity.} The five evaluation criteria were defined by the same author who designed the artefact, so they measure what the framework was built to do. This risks a benchmark that cannot fail. The mitigation is the pair of external reference points in Section~\ref{sec:results}: NIST's published informative references and the OpenSSF Scorecard output, neither of which was authored here and neither of which was adjusted after the results were seen.

\textit{Internal validity.} Rules, evidence, and ground truth share one author, so agreement between the engine and the ground truth partly reflects shared assumptions rather than independent confirmation. The adversarial probes were introduced specifically to attack the engine from outside those assumptions, and they found two defects that the criteria had not surfaced. That the probes succeeded is itself evidence that the main evaluation was not adversarial enough on its own.

\textit{External validity.} The scope is 26 requirements from four standards, one simulated organisation, one sector. Nothing here supports a claim about full standard coverage, about production evidence at scale, or about sectors with different evidentiary conventions. The synthetic case study also contains gaps that were placed deliberately, so gap detection demonstrates that the mechanism works and not that it would find gaps nobody had thought to introduce.

\textit{Conclusion validity.} Eleven determinations is far too few for statistical inference. The paper therefore reports counts and proportions, makes no significance claims, and attaches no confidence intervals to figures such as 88.5\,\% or 57.1\,\%; these describe the sample and nothing beyond it. Reliability is the one area where the design is strong: the rule engine is deterministic, so the reported results reproduce exactly from the accompanying artefact. The AI-assisted mapping component is not deterministic and its behaviour is reported separately for that reason.

\subsection{Future Work}
Four directions follow directly from the limitations. First, evaluation on real, multi-organisation data with independent annotators would establish ecological validity and inter-rater agreement. Second, the requirement set and standards coverage can be widened, including AI-specific obligations such as the EU AI Act \cite{eu_ai_act}. Third, evidence authentication would close the declared-versus-substantiated gap identified by the adversarial probes. Fourth, the AI-assisted mapping component can be benchmarked at scale once provider access is stable, reporting accepted-mapping stability across repeated runs.

% ============================================================
\section{Conclusion}
This paper presented AspisAI, a canonical, machine-interpretable governance framework for automated multi-standard compliance monitoring. Evaluated on a simulated critical-infrastructure case study, the prototype demonstrated high coverage, accurate and traceable compliance decisions, and reliable gap detection within its scope. Future work includes evaluation on real organisational evidence, expansion of the requirement set and additional frameworks (e.g.\ NIS2, the NCSC CAF, IEC~62443), richer rule formalisms, and integration into CI/CD toolchains.

% ============================================================
\section*{Acknowledgment}
This work was supported partially by the European Union in the framework of ERASMUS MUNDUS, Project CyberMACS (Project \#101082683) (\url{https://cybermacs.eu}). The first author thanks Franziska Schwarz for her guidance and support during this research.

% ============================================================
\printbibliography

\end{document}